Kinetics of the Free-Radical Polymerization of Isobornyl Methacrylate in the Presence of Polyisobutylenes of Different Molar Masses

**Ezequiel R. Soulé, Julio Borrajo, and Roberto J. J. Williams***

*Institute of Materials Science and Technology (INTEMA), University of Mar del Plata and National Research Council (CONICET), J. B. Justo 4302, 7600 Mar del Plata, Argentina*

*To whom correspondence should be addressed. E-mail: williams@fi.mdp.edu.ar

ABSTRACT: The scope of this study was to investigate the effect of a linear polymer dissolved in a reactive monomer on the kinetics of free-radical polymerization before the start of phase separation. The selected system was a solution of polyisobutylene (PIB) in isobornyl methacrylate (IBoMA), polymerized at 80 ºC in the presence of benzoyl peroxide (BPO). A ternary phase diagram of PIB, IBoMA and poly(isobornyl methacrylate) (PIBoMA), was built at 80 ºC both employing physical blends or determining the phase separation conversion in the course of polymerization. Cloud-point curves (CPC) obtained by both methods were coincident within experimental error. They were shifted to lower conversions when increasing the molar mass of PIB. Different PIBs exhibiting CPC at advanced conversions were selected for the kinetic study performed employing differential scanning calorimetry (DSC) at 80 ºC. A simple kinetic model for free-radical polymerizations describing the relevant termination rate constant in terms of the free-volume theory, provided a consistent fitting of the polymerization rates in the conversion range where the solution remained homogeneous. Increasing the molar mass of PIB led to an increase in polymerization rate due to the decrease in free volume and the corresponding decrease of the termination rate. Increasing the amount of a particular PIB in the initial formulation led to a less marked gel effect, explained by the smaller relative variation of free volume with conversion. The dimensionless free volume of PIB obtained from the kinetic model was found to increase with the volume concentration of chain ends, as expected. Under conditions where phase separation took place at very low conversions, the overall polymerization rate exhibited the presence of two maxima (gel effects), representing the polymerization in two different phases. The first maximum was associated to the polymerization taking place in the phase lean in PIB and the second maximum was associated to the polymerization of the monomer that was initially fractionated with PIB.

**Introduction**

An initial homogeneous solution of a polymer in a monomer of different chemical nature usually becomes phase separated upon polymerization of the monomer. The main driving force of this polymerization-induced phase separation process is the decrease in the absolute value of the contribution of the entropy of mixing to the Gibbs free energy.[1] The particular case of a linear polymer dissolved in a monomer that generates another linear polymer through a free-radical polymerization, is represented by the synthesis of high-impact polystyrene (HIPS) starting from a solution of polybutadiene in styrene. At a few percent conversion the system becomes phase separated because the two polymer solutions are incompatible.[2,3] Examples of similar systems that have been described in the literature are solutions of polyethylene in different solvents such as butyl methacrylate,[4] styrene,[5,6] and isobornyl methacrylate,[7] solutions of polystyrene in 2-chlorostyrene,[8,9] and solutions of poly(dimethylsiloxane-*co*-diphenylsiloxane) in 4-chlorostyrene.[10] In some of these systems the linear polymer is added to improve properties of the other linear polymer generated by polymerization. This is the case of HIPS where polybutadiene is added to generate a dispersed rubbery phase that improves the toughness of polystyrene. In other systems the linear polymer added to the initial formulation is the desired product and the monomer acts as a reactive solvent that facilitates its processing. Upon polymerization there is a phase inversion leading to a matrix formed by the initial linear polymer and dispersed domains constituted by the polymer formed by the reactive solvent. This process may be used to facilitate processing of polyethylene or polystyrene, e.g., to produce a correct filling of parts with a complex geometry.

Most of the studies in the field are focused on aspects related to the phase separation process, morphologies generated and properties of the resulting materials. The structures formed in the course of phase separation are determined by the competition between the polymerization and the phase separation rates. Therefore, the way in which the presence of the linear polymer affects the kinetics of the free-radical polymerization is important for these phase separation studies but has not been given a proper consideration in the literature. The analysis is complex by the fact that phase separation usually occurs at very low conversions. Therefore, the polymerization takes place in a single phase only in a small conversion range and, after phase separation it takes place in two (or more) phases exhibiting different compositions.

The aim of this study was to analyze the way in which a mechanistic kinetic model can take into account the effect of the linear polymer on the free-radical polymerization rate. This is relevant for a kinetic description in the initial solution before phase separation, and in each one of the individual phases generated after phase separation.

We found a particular system consisting of polyisobutylene (PIB) dissolved in isobornyl methacrylate (IBoMA), in which the cloud-point conversion could be varied in a broad range by using PIB oligomers of different molar masses. In this way, the kinetic analysis in the presence of PIB could be extended to almost the whole conversion range. A kinetic model is proposed where the effect of PIB is explicitly taken into account in the expression of the chain termination rate. The consistency of the kinetic model to describe the polymerization rate under conditions where the gel effect is either very important or almost insignificant will be discussed. In a final section experimental results of the overall polymerization rate after phase separation will be analyzed in a qualitative way.

As has been recently recognized,[11-13] many kinetic studies reported in the literature for free-radical polymerizations are based on experimental results obtained under non-isothermal conditions due to the lack of temperature control when the polymerization becomes self-accelerated (gel effect). A convenient technique to generate isothermal kinetic data is differential scanning calorimetry (DSC), due to the small masses used and the corresponding high heat transfer rates. Besides, the signal is directly proportional to the reaction rate, a fact that is important for kinetic analysis. DSC was the technique selected in this study to obtain the polymerization kinetics. Pans were sealed in an inert atmosphere avoiding the presence of oxygen that produces a significant retardation effect on the polymerization rate.

**Kinetic Model**

In order to minimize the use of adjustable parameters, a simple kinetic model was used to fit the experimental curve of polymerization rate as a function of conversion. The model was previously used to analyze the free-radical polymerization of pure isobornyl methacrylate (IBoMA),[14] and is now extended to describe the reaction in the presence of polyisobutylene (PIB).

Due to the volume contraction in the course of polymerization, the instantaneous volume ($V$) of the system varies with monomer conversion ($x$) according to:

$$(V/V^0) = [V^0_{IBoMA}(1+\varepsilon x) + V_{PIB}]/[V^0_{IBoMA} + V_{PIB}] = (1+\varepsilon x + \phi)/(1+\phi) \tag{1}$$

where $V^0 = V_{PIB} + V^0_{IBoMA}$, is the total initial volume, $\phi = V_{PIB}/V^0_{IBoMA}$, $\varepsilon = (\rho_m - \rho_p)/\rho_p$, is the volume expansion factor (a negative value), and $\rho_m$ and $\rho_p$ are, respectively, the densities of the monomer and polymer (PIBoMA).

By calling $I$, $R$ and $M$ the number of moles of initiator, free radicals and monomer, respectively, the rate equations may be written as follows:

$$(1/V) \, dI/dt = - k_d \, (I/V) \tag{2}$$

$$(1/V) \, dR/dt = 2f \, k_d \, (I/V) - 2k_t \, (R/V)^2 \tag{3}$$

$$(1/V) \, dM/dt = - k_p \, (R/V) \, (M/V) \tag{4}$$

where $k_d$, $k_t$ and $k_p$ are, respectively, the specific rate constants for the initiator decomposition, termination and propagation; $f$ is the efficiency of the initiation process. Solving eq 2 and replacing the solution in eq 3, and using eq 1 and the definition of conversion, $x = 1 - M/M_0$, the following set of kinetic equations may be written:

$$d(R/V^0)/dt = 2f k_d (I_0/V^0) \exp(-k_d t) - 2k_t (R/V^0)^2 (1+\phi)/(1+\varepsilon x +\phi) \quad (5)$$

$$dx/dt = k_p (1-x) (R/V^0) (1+\phi)/(1+\varepsilon x +\phi) \quad (6)$$

Both $k_t$ and $f$ have to be considered a function of conversion. The usual finding is that $f$ remains almost constant up to intermediate conversion and only then exhibits a sharp drop.[15,16] Experimental values of polymerization rates were fitted with constant values of $k_d$, $k_p$ and $f$ up to intermediate conversions. At high conversions the value of $f$ was continuously adjusted to fit the experimental reaction rate.

In a previous paper,[14] we described the kinetics for the pure monomer with a similar set of kinetic equations (with $\phi = 0$), and assuming that termination was controlled by translational diffusion up to very high conversions. It was not necessary to introduce a segmental diffusion term at the beginning of polymerization and termination by reaction-diffusion was estimated to be negligible at least up to conversions close to 70 %. A similar approach will be used here. Translational diffusion is expressed in terms of the Fujita-Doolittle theory based on the free volume:[17-19]

$$D = D_0 \exp(-\alpha_0 /V_f) \quad (7)$$

where $\alpha_0$ is a constant and $V_f$ is the total free volume given by the product of the free volume and the volume fraction of every component in the solution:

$$V_f = [V_{f,\text{IBoMA}} (1-x) + V_{f,\text{PIBoMA}} x (1+\varepsilon) + V_{f,\text{PIB}} \phi]/(1+\varepsilon x +\phi) \quad (8)$$

The free volumes of every component depend only on temperature and may be taken as constants for an isothermal polymerization.

The specific rate for termination by translational diffusion, $k_t$, may be considered proportional to $D$.

$$\log k_t = \alpha_1 - \alpha_0/(2.303 V_f) \quad (9)$$

where $\alpha_1$ is a constant.

Adding and substracting $\alpha_0/(2.303\ V_{f,PIBoMA})$, leads to:

$$\log k_t = \beta_1 + (\alpha_0/2.303)(1/V_{f,PIBoMA} - 1/V_f) \tag{10}$$

where $\beta_1 = \alpha_1 - \alpha_0/(2.303\ V_{f,PIBoMA})$.

Eq 10 may be rearranged to give:

$$\log k_t = \beta_1 + [1 - x + \beta_2\phi]/[\beta_3(1+\varepsilon x+\phi) + \beta_4(1 - x) + \beta_2\beta_4\phi] \tag{11}$$

where

$$\beta_2 = (V_{f,PIB} - V_{f,PIBoMA})/(V_{f,IBoMA} - V_{f,PIBoMA})$$

$$\beta_3 = (2.303/\alpha_0)(V_{f,PIBoMA})^2/(V_{f,IBoMA} - V_{f,PIBoMA})$$

$$\beta_4 = (2.303/\alpha_0)(V_{f,PIBoMA})$$

Eq 11 is basically based on Doolittle's free volume model and on the hypotheses that $\alpha_0$ is the same for every component and free volumes are additive (although this last hypothesis may be questioned).[20] The grouping of parameters enabled to reduce the number of adjustable parameters from five ($\alpha_0$, $\alpha_1$ and the three free volumes) to four ($\beta_1$ to $\beta_4$). Although the concept of free volume is useful to analyze the dependence of molecular mobility on temperature or composition, it is not possible to assess its value from theoretical arguments mainly due to the fact that the magnitude of the occupied volume remains a matter of conjecture and can be estimated only indirectly.[20] Therefore, we kept $\beta_1$ to $\beta_4$ as adjustable parameters. Although at a first impression it may be considered that this gives a high versatility to fit almost any kinetic curve, the following should be taken into account: a) different kinetic curves obtained by varying the initial amount of a particular PIB should be fitted using the same set of values of the four adjustable parameters, b) kinetic curves obtained using PIB of different molar masses should be fitted using the same values of $\beta_1$, $\beta_3$ and $\beta_4$ and keeping only $\beta_2$ as an adjustable parameter ($V_{f,PIB}$ that depends on the molar mass of PIB is only present in this parameter), c) fitted $\beta_2$ values must be such that $V_{f,PIB}$ increases when its molar mass decreases. Therefore, experimental results will provide a strong test of the suitability of the proposed kinetic model.

**Experimental Section**

**Materials.** Isobornyl methacrylate (IBoMA, Aldrich) was used as received. It contained 150 ppm of *p*-methoxyphenol (MEHQ, methyl ether hydroquinone) as inhibitor. Benzoyl peroxide (BPO, Akzo-Nobel) was used as initiator. A series of polyisobutylene (PIB) oligomers of different molar masses was available from Repsol YPF (Argentina).[21] The denomination, molar mass averages determined by SEC, and densities at 15 ºC of the family of selected polyisobutylenes is shown in Table 1.

For initial studies of the miscibility of PIBs in mixtures of IBoMA and its polymer (PIBoMA), it was necessary to synthesize the polymer using the same experimental conditions as those used in the kinetic study. PIBoMA was synthesized by heating about 2 g of an IBoMA –BPO solution containing 2 wt % BPO, placed in a closed glass tube at 80 ºC for 1 h. The polymerization was completed in 30 min at 140 ºC (no residual reaction heat was observed by differential scanning calorimetry). Average molar masses of PIBoMA determined by SEC were: $M_n = 1.60 \times 10^5$ and $M_w = 9.35 \times 10^5$.

**Molar Mass Distributions.** Molar mass distributions were determined by size exclusion chromatography (SEC), using solutions containing 2-8 mg of polymer per mL of tetrahydrofuran (THF). A Knauer K-501 device provided with a refractive index detector (K-2301) and a set composed of one ultrastyragel $10^4$ column and three styragel HR3, HR1 and HR0.5 columns (Waters), was employed. THF was used as a carrier at 1 mL/min. Molar mass distributions were obtained using a universal calibration curve determined with polystyrene standards and the Mark-Houwink constants of PIB[22] and PIBoMA[23] in THF. In the case of PIB these constants are reported for a range of low molar masses that was not specified.[22]

**Miscibility Curves in PIB-IBoMA-PIBoMA Blends.** Cloud-point curves of ternary solutions at 80 ºC were determined both in physical blends of the three components and in the course of a polymerization.

Cloud-point curves of physical blends were obtained as follows. First a blend of desired composition was prepared using dichloromethane to facilitate the mixing process, followed by solvent evaporation at about 40 ºC, until the theoretical constant weight was obtained (the vapor pressure of IBoMA at this temperature is negligible). The blend was then placed between two glass slides separated by a 0.5 mm stainless steel spacer and the cloud-point temperature was determined using a Leica DMLB microscope provided with a video camera (Leica DC100) and a hot stage (Linkam THMS 600). Samples were heated to a temperate above the cloud-point curve and then cooled at 1 ºC/min until the cloud-point was observed. The procedure was performed several times observing good repeatability of the cloud-point temperature. To obtain the particular composition exhibiting a cloud-point temperature of 80 ºC, blends containing the same ratio of two of the components and a variable amount of the third component were studied. A plot of cloud-point temperatures vs. the amount of the third component was generated. Interpolation at 80 ºC enabled to obtain the desired composition. The procedure was repeated by varying the initial fixed ratio of two components. In this way, cloud-point curves at 80 ºC were generated.

The cloud-point of PIB-IBoMA solutions containing 2 wt % BPO with respect to the monomer, was determined in the course of polymerizations carried out in the cell of a UV-visible spectrophotometer (Shimadzu UV-1601PC), kept at 80 ºC with water circulating from a thermostat. Samples were placed in a 2-3 mm gap between two glass windows located in a metallic frame that fitted exactly into the cell of the spectrophotometer. Exploratory runs were performed placing a thermocouple inside the sample and confirming that polymerization was conducted under isothermal conditions. The intensity of light at $\lambda = 330$ nm, transmitted through the sample in the course of polymerization, was continuously monitored. The cloud-point was defined at the onset of the decrease in the light intensity. At this time, the sample was rapidly cooled in a water-ice mixture. Partially converted samples were dissolved in methylene chloride (3.5 g in 100 g solvent), and Fourier transformed infrared spectra were recorded (FTIR, Genesis II, Mattson, liquid cell with NaCl windows and a 0.2 mm Teflon spacer). The monomer conversion was followed by measuring the absorbance of the C=C stretching vibration at 1640 cm$^{-1}$. To take into account small variations in the initial concentration, the band at 1455 cm$^{-1}$ assigned to a combination of asymmetrical C-CH$_3$ vibrations and to C-H bending in CH$_2$ groups was used as a reference. By calling $h = A_{1640}/A_{1455}$, the monomer conversion was defined as

$$x = 1 - [(h(t) - h_{PIBoMA})/(h(0) - h_{PIBoMA})] \tag{12}$$

where $h_{PIBoMA}$ takes into account the residual (very small) absorbance at 1640 cm$^{-1}$ present in the polymer.

**Kinetics.** Polymerization of PIB-IBoMA solutions containing 2 wt % BPO (expressed per mass of BPO + IBoMA), were polymerized at 80 °C in the cell of a differential scanning calorimeter (DSC, Pyris 1, Perkin-Elmer). DSC pans containing the sample were sealed in a closed chamber under a nitrogen flow to eliminate air in contact with the sample surface (the presence of oxygen causes a significant retardation effect). Conversion and polymerization rate were defined as:

$$x = \Delta H(t)/\Delta H_T \tag{13}$$

$$dx/dt = (1/\Delta H_T)(dH/dt) \tag{14}$$

where $\Delta H_T$ was calculated on the basis of the total heat of reaction of pure IBoMA, equal to 54.2 kJ/mol.[14] Some DSC runs of PIB-IBoMA solutions were performed under dynamic conditions at 5 ºC/min, confirming the value of the total reaction heat within experimental error.

**Results and Discussion**

**Cloud-Point Curves of PIB-IBoMA-PIBoMA Blends at 80 ºC.** Figure 1 shows the miscibility of three different PIBs determined both in physical blends and in the course of a polymerization-induced phase separation. Solutions were homogeneous above the respective cloud-point curves. The whole set of PIBs was miscible with IBoMA at 80 ºC.

While PIB025 remains miscible up to very high conversions, PIB5 phase separates at intermediate conversions and PIB30 exhibits a much lower miscibility. A 50 wt % solution of PIB30 phase separates at the start of the polymerization. Therefore, the conversion range in which the kinetic study in a homogeneous solution could be performed decreased when increasing the molar mass of PIB. It is interesting to observe that small variations of the average molar masses of PIB led to a significant shift of the cloud-point curve. This arises from the high sensitivity of the entropic contribution to the free energy of mixing in the range of low molar masses.[1]

Within the experimental error of these determinations, cloud-point curves obtained in physical blends or in the course of a polymerization were the same. Similar results were recently reported by Liskova and Berghmans for the polymerization of styrene in the presence of polyethylene wax.[6]

**Grafting of PIBoMA on PIB During Polymerization.** The possible grafting of PIBoMA on PIB through chain transfer reactions produced in the course of polymerization was also investigated. Grafting involves the termination of a propagating chain of PIBoMA by abstraction of a hydrogen radical from the PIB backbone. The generated radical in the PIB structure will re-initiate a new PIBoMA chain. It is assumed that re-initiation takes place at the same rate than propagation so that this event has no effect on the polymerization kinetics.

SEC chromatograms of physical blends of PIB-PIBoMA containing 50 wt % PIB were compared with those of blends of the same composition produced after complete polymerization of PIB-IBoMA solutions at 80 °C (Figure 2). The peak of PIBoMA is present at low elution times and the peak of PIB appears at high elution times.

In order to quantify the fraction of grafted PIB the area under both peaks in the SEC chromatogram was obtained by deconvolution of the spectra using Gaussian components. This led to the following ratio of areas under both peaks: $(A_{PIB}/A_{PIBoMA})_{phys}$ and $(A_{PIB}/A_{PIBoMA})_{chem}$. By calling $A_{PIB}(g)$, the fraction of PIB grafted to PIBoMA chains, the following balances may be written:

$$(A_{PIB})_{chem} = (A_{PIB})_{phys} - A_{PIB}(g) \tag{15}$$

$$(A_{PIBoMA})_{chem} = (A_{PIBoMA})_{phys} + A_{PIB}(g) \tag{16}$$

From eqs 15 and 16, the mass fraction of grafted PIB, $w_{PIB}(g)$, may be calculated:

$$w_{PIB}(g) = A_{PIB}(g)/(A_{PIB})_{phys} = [1 - (A_{PIB}/A_{PIBoMA})_{chem}/(A_{PIB}/A_{PIBoMA})_{phys}] /$$
$$[1 + (A_{PIB}/A_{PIBoMA})_{chem}] \tag{17}$$

The resulting mass fractions of grafted PIB were 0.12 for PIB025, 0.09 for PIB5, and 0.06 for PIB150. Grafting was not very significant for any of these blends taking into account that initial formulations contain 50 wt % PIB. Grafting was more important for the more miscible PIB as could be expected.

**Polymerization Kinetics.** PIB-IBoMA solutions containing 2 wt % BPO with respect to IBoMA + BPO, were polymerized at 80 °C in the DSC. An induction period explained by the presence of inhibitors was observed for every blend. This was in fact convenient because it enabled to have a defined baseline at 80 °C before the start of polymerization.

Figure 3 shows the polymerization rate as a function of time (Figure 3a) or conversion (Figure 3b), for solutions of PIB025 in IBoMA. For this particular blend phase separation takes place at very high conversions. Therefore, the influence of PIB on kinetics could be investigated in practically the whole conversion range. The presence of a maximum in reaction rate, the so-called gel or Tromsdorff effect, became less significant as the PIB concentration increased and practically disappeared for the blend with 50 wt % PIB. For the neat monomer, the polymerization rate exhibited a sharp decrease at a conversion close to 0.8 due to vitrification.[14] In blends with PIB the final conversion was higher due to the plasticizing effect of residual PIB dissolved in PIBoMA.

In order to fit the kinetic model the following values of specific rate constants and parameters were taken from the literature:[14,23] $k_d = 4.17 \times 10^{-5}$ s$^{-1}$, $k_p = 2 \times 10^3$ L mol$^{-1}$ s$^{-1}$, $f = 0.8$ (up to a conversion close to the gel effect), $\varepsilon = -0.0755$. The value of $\phi$ for a blend containing a particular wt% PIB was calculated assuming that there was no volume change upon mixing:

$$\phi = V_{PIB}/V^0_{IBoMA} = [\text{wt\%}/(100 - \text{wt\%})](\rho_{IBoMA}/\rho_{PIB}) \tag{18}$$

Values of densities of different PIBs at 15 °C are reported in Table 1. The density of the monomer at the same temperature was: $\rho_{IBoMA} = 0.983$ g/cm$^3$. It was assumed that the ratio of densities did not vary with temperature.

The initial concentration of initiator was given by

$$(I_0/V^0) = 0.077/(1+\phi) \text{ mol L}^{-1} \tag{19}$$

The only remaining rate constant of the set of kinetic equations, eqs 5 and 6, was the termination constant $k_t$, expressed by eq 11 in terms of four fitting parameters: $\beta_1$ to $\beta_4$. The best set of these four parameters that could fit the four experimental curves up to conversions close to 0.45, was searched using the Levenberg-Marquardt algorithm included in Mathcad

2001 Professional. The following set was obtained: $\beta_1 = 2.482$, $\beta_2 = 0.799$, $\beta_3 = 0.145$, $\beta_4 = 0.0522$. The resulting fitting is shown by the full lines plotted in Figure 3b.

The fitting of kinetic curves was extended to conversions close to the start of vitrification, letting the efficiency of initiator decomposition ($f$) to decrease in order to adjust the experimental curves. Figure 4 shows the variation of $f$ with conversion for every one of the blends. Obviously, the reliability of the kinetic model is limited to the range of conversions where the value of $f$ was kept constant (up to conversions close to 0.45). However, it is important to realize that monomer depletion alone cannot be taken as responsible of the significant decrease in polymerization rate observed above intermediate conversions.

Figure 5 shows the variation of the termination rate constant with conversion for PIB025-IBoMA blends containing different wt % PIB. Increasing the PIB content produced a decrease in the initial value of $k_t$ due to the smaller value of the free volume of PIB025 with respect to the one of IBoMA. However, as the free volume associated with PIB remained constant, the total free volume decreased less with conversion when increasing the PIB content of the blend. This led to a slower decrease of $k_t$ with conversion. And in turn, this explains the slower increase of the polymerization rate with conversion observed when increasing the wt % PIB.

Two factors affected the initial values of polymerization rates when varying the PIB wt % in the blend. Increasing the wt % PIB produced a decrease in both the initiator concentration (eq 19) and in the termination rate constant (Figure 5). While the first factor leads to a decrease in the polymerization rate, the second one increases the reaction rate. A compensation effect between both factors led to an initial polymerization rate that was practically independent of the wt % PIB in the blend.

The polymerization kinetics of blends of PIB5-IBoMA was also fitted with the kinetic model up to cloud-point conversions (or to a conversion close to the maximum if phase separation occurred after this point). Figure 6 shows the fitting obtained keeping the same values for $\beta_1$, $\beta_3$ and $\beta_4$ and searching for the best value of $\beta_2$ (this parameter is the only one that depends on the free volume of PIB). The best value of this parameter was $\beta_2 = 0.733$ which is lower than the one found for PIB025 ($\beta_2 = 0.799$). As $\beta_2$ increases with $V_{f,PIB}$ it is expected that its value should decrease when increasing the molar mass of PIB as was indeed the case.

In order to find values of $\beta_2$ for PIBs of higher molar masses, kinetic curves obtained for PIB10-IBoMA and PIB30-IBoMA blends, were fitted in the initial conversion range before the start of phase separation. Corresponding values found for $\beta_2$ were 0.619 and 0.538, respectively.

In order to translate these differences in $\beta_2$ values into differences in physical parameters, the dimensionless free volume associated with a particular PIB may be calculated:

$$V_{f,PIB}^{*} = (V_{f,PIB} / \alpha_0) = \beta_4(1 + \beta_2\beta_4/\beta_3)/2.303 \qquad (20)$$

The free volume associated with a particular PIB must be an increasing function of the concentration of chain ends which is proportional to $(\rho/M_n)$. Figure 7 shows the dimensionless free volume associated with PIB as a function of $(\rho/M_n)$. A rough linear correlation was obtained consistently with the physical meaning of the parameters of the kinetic model.

Another interesting feature of the polymerization kinetics results from the comparison of reaction rates observed for formulations containing the same amount of different PIBs. Figure 8 shows polymerization rates for solutions containing 15 wt % of PIB025 or PIB30, and 50 wt % of PIB 025 or PIB5. An increase in the molar mass of PIB produced an increase of the polymerization rate in the conversion region where both solutions remained homogeneous (using cloud-point data of Figure 1, these conversions are close to $x = 0.40$ for both cases). This is due to the decrease of the free volume contributed by the PIBs of higher molar mass and the consequent decrease of the termination rate.

**Overall Polymerization Rate in a Phase-Separated Blend.** A solution containing 50 wt % PIB30 in IBoMA phase separates at very low conversions (Figure 1). At this time two phases are generated, one rich in IBoMA/PIBoMA and the other one rich in PIB, with a fractionation of the initiator between both phases. The polymerization rate will continue with different rates in both phases. Mass transfer between them (or secondary phase separation processes as those leading to the generation of a salami structure in HIPS), will produce a continuous enrichment of one phase in PIBoMa and the other one in PIB.

Figure 9 shows the polymerization rates observed for IBoMA solutions containing 50 wt % PIB025 (homogeneous polymerization up to a conversion close to 0.90), and 50 wt % PIB30 (polymerization in a phase separated system from very low conversions). For the phase-separated blend two maxima in polymerization rate are observed, associated with the gel effect occurring at different times in both phases. Although the signal recorded at any time is the sum of the polymerization rate in both phases, the region around the first maximum must receive a significant contribution from the IBoMA/PIBoMA-rich phase while the region around the second maximum can be related to the polymerization of the monomer initially fractionated with the PIB-rich phase. The IBoMA/PIBoMA-rich phase exhibits a fast polymerization rate and a marked gel effect.

Figure 10 shows a comparison of polymerization rates of neat IBoMA and of the solution containing 50 wt % PIB30. Although initial polymerization rates (just before phase separation) were similar for both systems, the gel effect in the IBoMA-rich phase was observed at shorter times for the phase-separated system. This is explained by the fact that the conversion necessary to produce the gel effect in the IBoMA-rich phase was attained at a lower overall conversion than in neat IBoMa. However, the maximum polymerization rate in the phase-separated system was lower than in neat IBoMA. This may be explained by the presence of a fraction of PIB30 in the IBoMA-rich phase, producing a similar decrease in the maximum polymerization rate as the one shown in Figure 3. Fractionation of the initiator between both phases might also affect the polymerization rate.

The fact that after the cloud point a free-radical polymerization might continue at significantly different rates in both phases opens the possibility of decoupling individual kinetics from the overall DSC signal. This might be of interest for the modeling of the polymerization-induced phase separation in industrial processes such as the synthesis of HIPS and related blends.

**Conclusions**

A simple kinetic model for free-radical polymerizations describing the relevant termination rate constant in terms of the free-volume theory, provided a consistent fitting of the polymerization rates of solutions of an oligomer (PIB) in a monomer (IBoMA), in the conversion range where the solution remained homogeneous. The fitting required the use of four adjustable parameters appearing in the expression of the termination rate constant. Three of the four parameters were invariant with the PIB molar mass and its concentration in solution, as expected from the model. The remaining parameter varied with the PIB molar mass in a way consistent with the decrease in its contribution to free volume when increasing its molar mass.

Apart from the trivial dilution effect, the main influence of PIB on the polymerization kinetics was related to its action over the termination rate constant. For a particular PIB, an increase in its concentration produced a decrease in the initial free volume of the system (the contribution of PIB to the total free volume was less than the one of the monomer). In turn this produced a decrease in the termination rate constant and an increase in the initial polymerization rate (counterbalanced by a dilution effect when the amount of added initiator is defined with respect to the monomer and not to the total volume). Increasing the molar mass of PIB produced a decrease in free volume and in the termination rate constant and a consequent increase in the polymerization rate.

The polymerization rate increased continuously from the beginning of reaction up to a maximum value due to the continuous decrease of the termination rate. This is the origin of the so-called gel effect. The presence of a maximum was mainly produced by the sharp decrease of the initiation efficiency. The addition of PIB to the initial formulation reduced the relative variation of free volume with monomer conversion because the fraction of free volume supplied by PIB remained constant. Therefore, the variation of the termination rate with conversion was attenuated as well as the gel effect. The free-volume model provided a consistent description of the gel effect either for the polymerization of the pure monomer where a significant auto-acceleration was observed, and for the polymerization of formulations containing 50 wt % PIB where the gel effect was not significant.

Increasing the molar mass of PIB led to a decrease in the cloud-point conversion. Under conditions where phase separation took place at very low conversions, the overall polymerization rate exhibited the presence of two maxima (two gel effects), representing the polymerization in two different phases. The first maximum was associated to the polymerization taking place in the phase lean in PIB that consequently showed a marked gel effect. The second maximum was associated to the polymerization of the monomer that was initially fractionated with PIB. The possibility of generating experimental information that separates the polymerization kinetics taking place in both phases might be of interest for the modeling of phase separation induced by a free-radical polymerization.

These results are relevant to model free-radical polymerization rates in solutions that include a linear polymer different than the one formed by reaction. Several processes based on this type of formulations are of practical significance, the most relevant one being the synthesis of high-impact polystyrene. The kinetic model can be used for the initial homogeneous system up to the cloud-point conversion and then it can take into account the differences in polymerization rates in each one of the generated phases.

**Acknowledgment.** The financial support of the National Research Council (CONICET), the National Agency for the Promotion of Science and Technology (ANPCyT), the University of Mar del Plata, and Fundación Antorchas, Argentina, is gratefully acknowledged.

**Table 1. Denomination, Molar Mass Averages and Densities of Selected Polyisobutylenes**

| Polyisobutylene | $M_n$ | $M_w$ | $\rho_{15°C}$ (g/cm$^3$) |
|---|---|---|---|
| PIB025 | 720 | 1316 | 0.869 |
| PIB5 | 916 | 2599 | 0.888 |
| PIB10 | 1180 | 2650 | 0.898 |
| PIB30 | 1557 | 3929 | 0.904 |
| PIB150 | 2460 | 6836 | 0.903 |

**Legends to the Figures**

**Figure 1.** Cloud-point curves of PIB-IBoMA-PIBoMA solutions at 80 °C, in mass fraction coordinates. PIB30, PIB5 and PIB025 are respectively represented by squares, triangles and circles. Filled symbols are experimental points of physical blends while unfilled symbols represent values obtained during polymerization.

**Figure 2.** Comparison of SEC chromatograms obtained for physical mixtures of PIB and PIBoMA containing 50 wt % PIB, and blends of similar composition obtained by polymerization of PIB solutions in IBoMA, at 80 °C. (a) PIB025, (b) PIB5, (c) PIB150.

**Figure 3.** Polymerization rate of PIB025-IBoMA solutions containing 0, 15, 30 and 50 wt % PIB025; (a) as a function of time, (b) as a function of conversion (symbols represent experimental values and full lines indicate the best fitting arising from the kinetic model).

**Figure 4.** Efficiency of the initiator decomposition as a function of conversion for PIB025-IBoMA blends containing different wt % PIB.

**Figure 5.** Termination rate constant as a function of conversion for PIB025-IBoMA blends containing different wt % PIB.

**Figure 6.** Polymerization rate of PIB5-IBoMA solutions containing 15, 30 and 50 wt % PIB5, plotted as a function of conversion (symbols represent experimental values and full lines indicate the best fitting arising from the kinetic model).

**Figure 7.** Dimensionless free volume associated with PIB as a function of $\Box(\rho/M_n)$.

**Figure 8.** Comparison of polymerization rates for solutions containing the same amount of different PIBs. (a)15 wt % PIB025 and PIB30, (b) 50 wt % PIB025 and PIB5.

**Figure 9.** Polymerization rate as a function of time for IBoMA solutions containing 50 wt % PIB025 or 50 wt % PIB30.

**Figure 10.** Comparison of polymerization rates for neat IBoMA and for a solution containing 50 wt % PIB30.

**Figure 1**

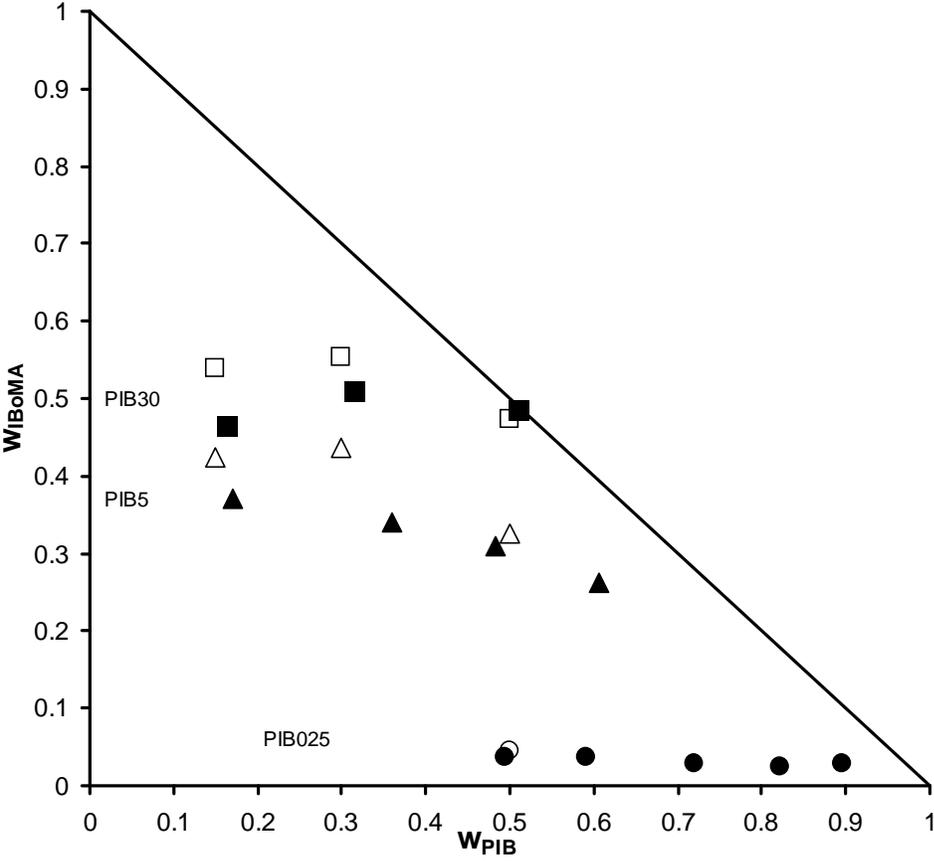

**Figure 2**

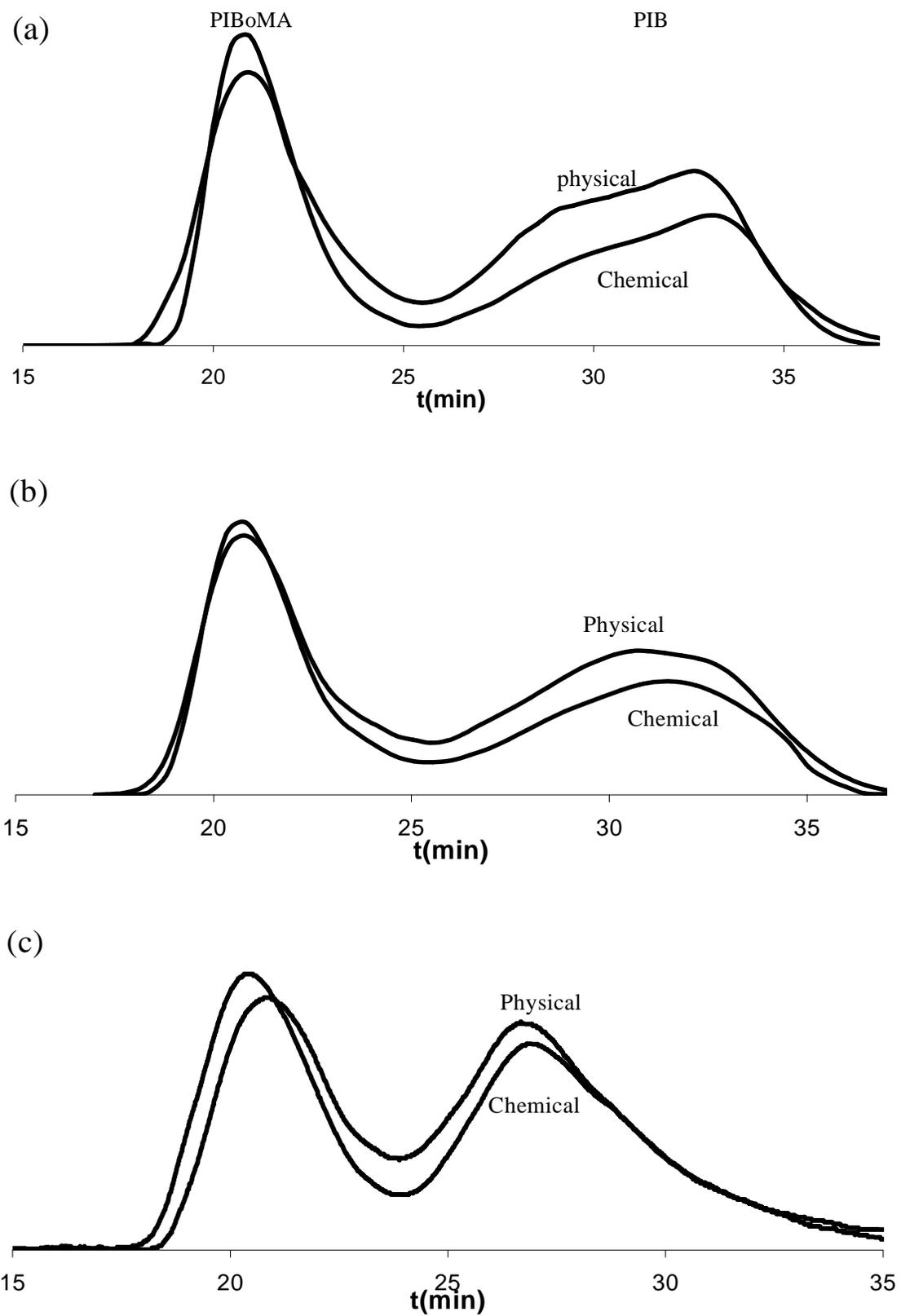

**Figure 3**

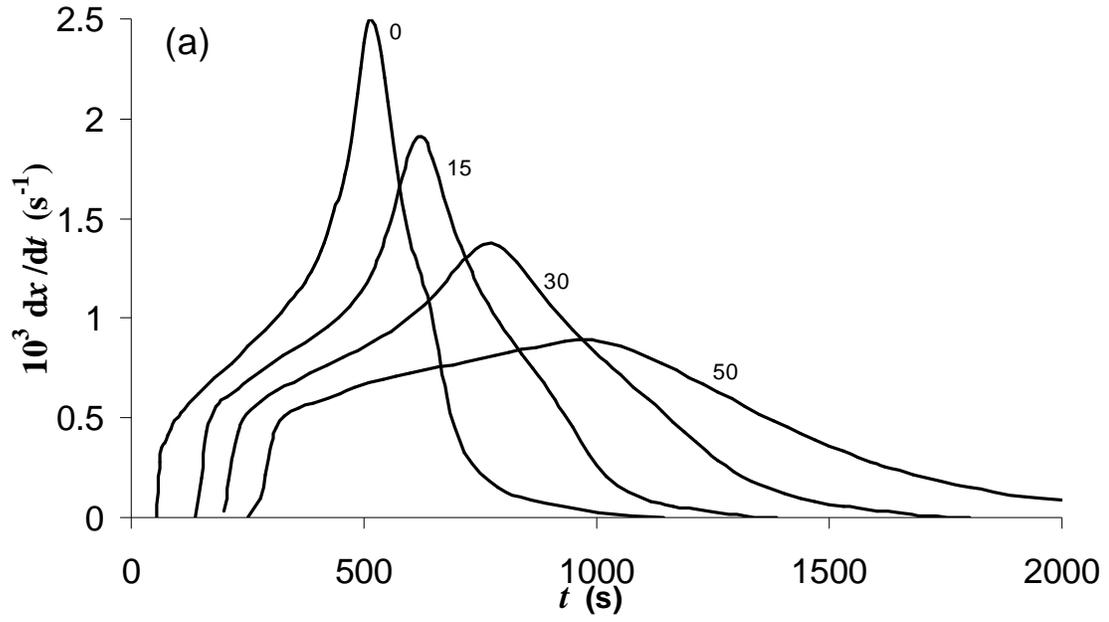

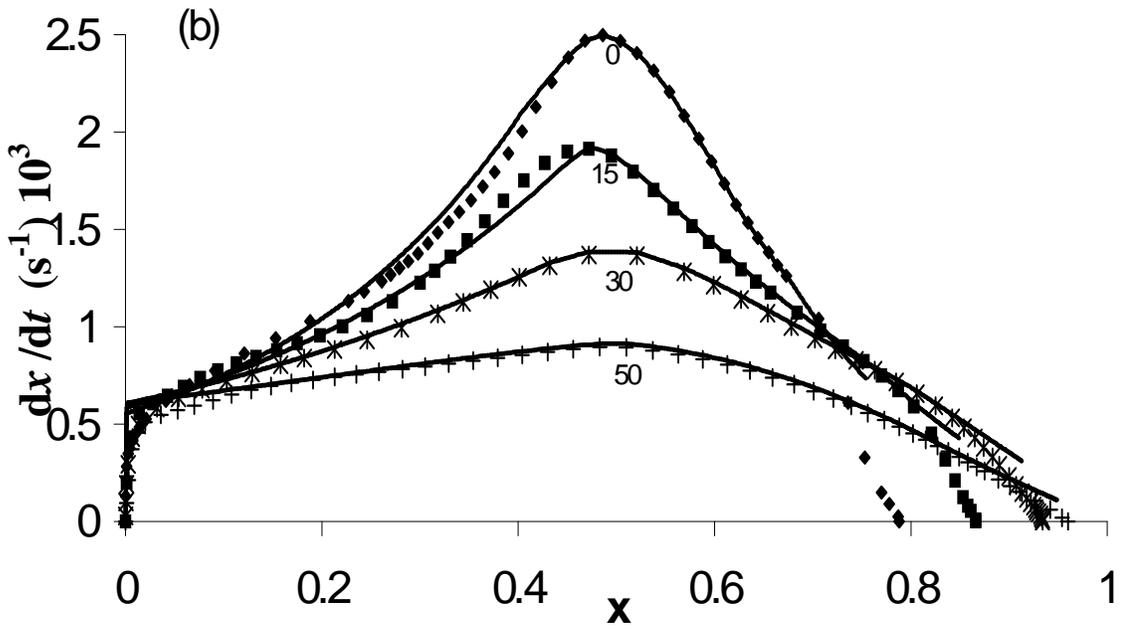

**Figure 4**

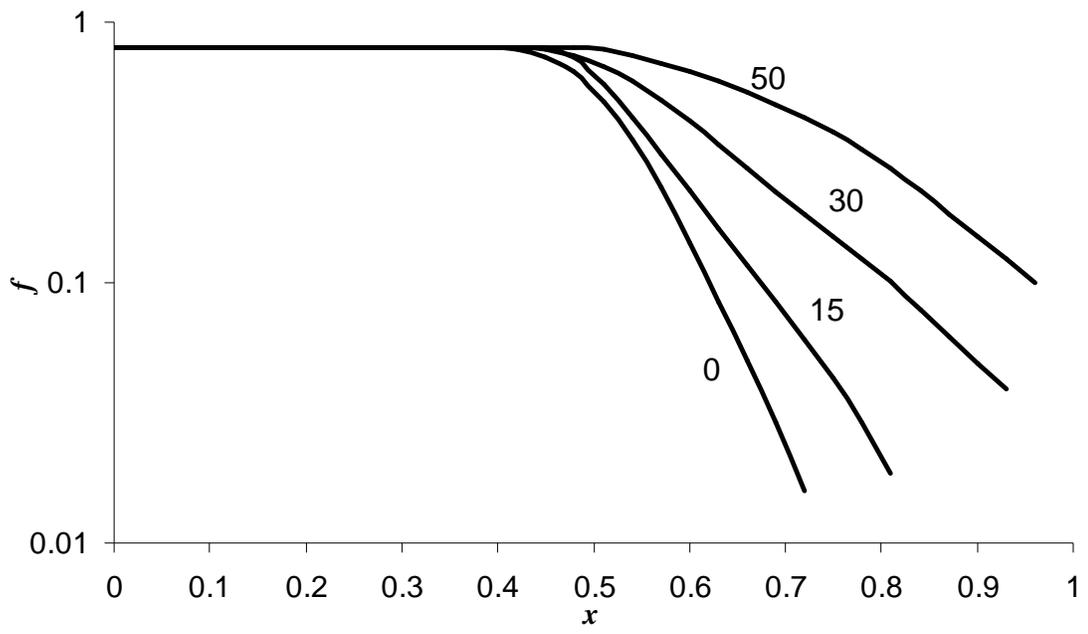

**Figure 5**

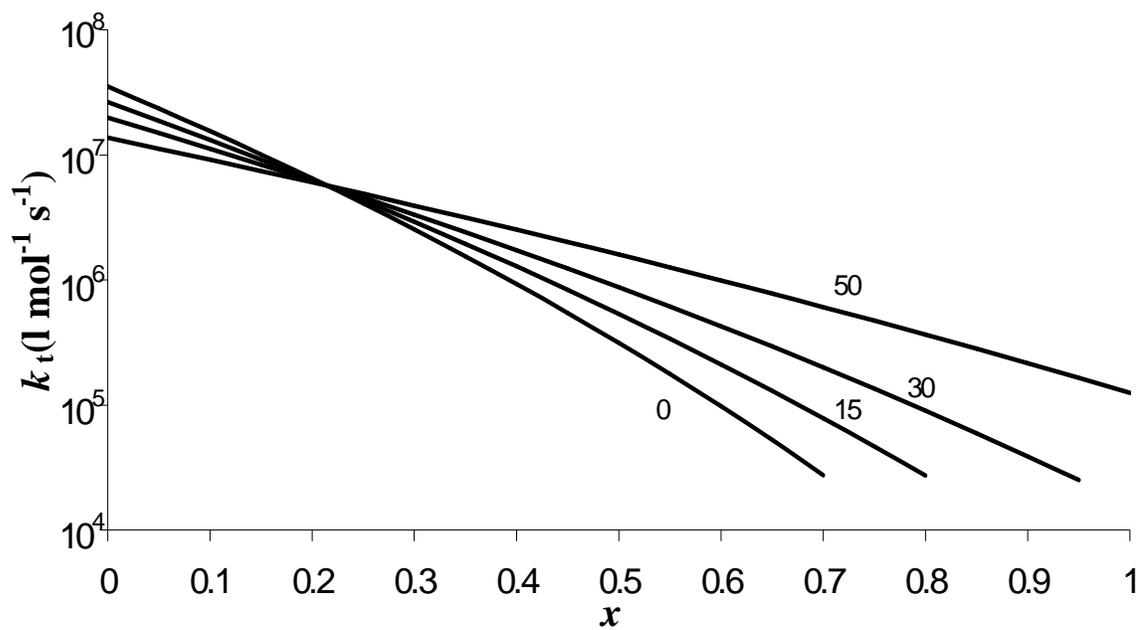

**Figure 6**

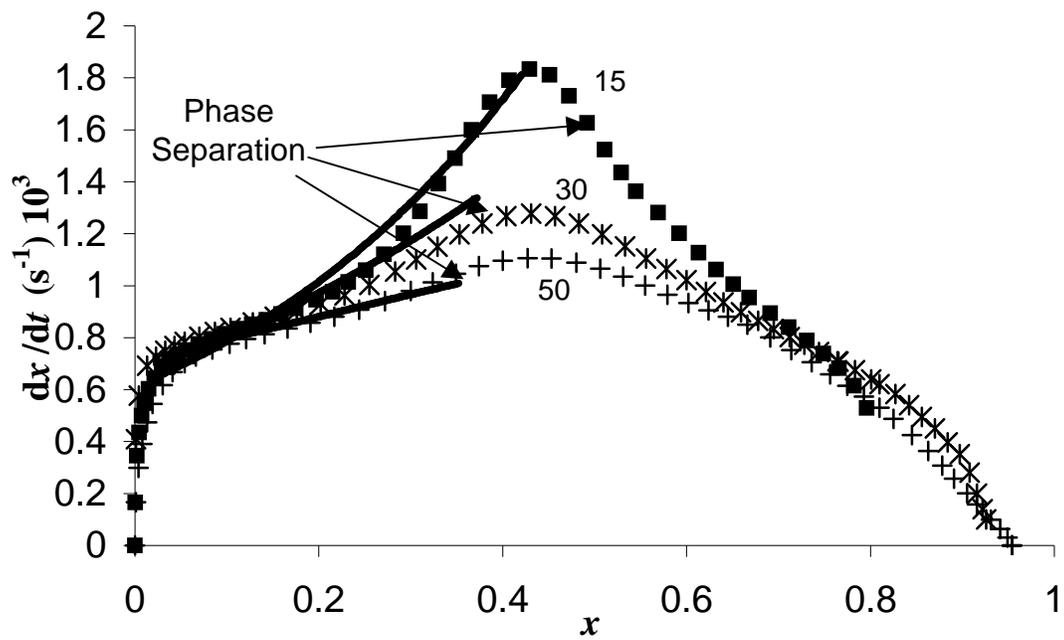

**Figure 7**

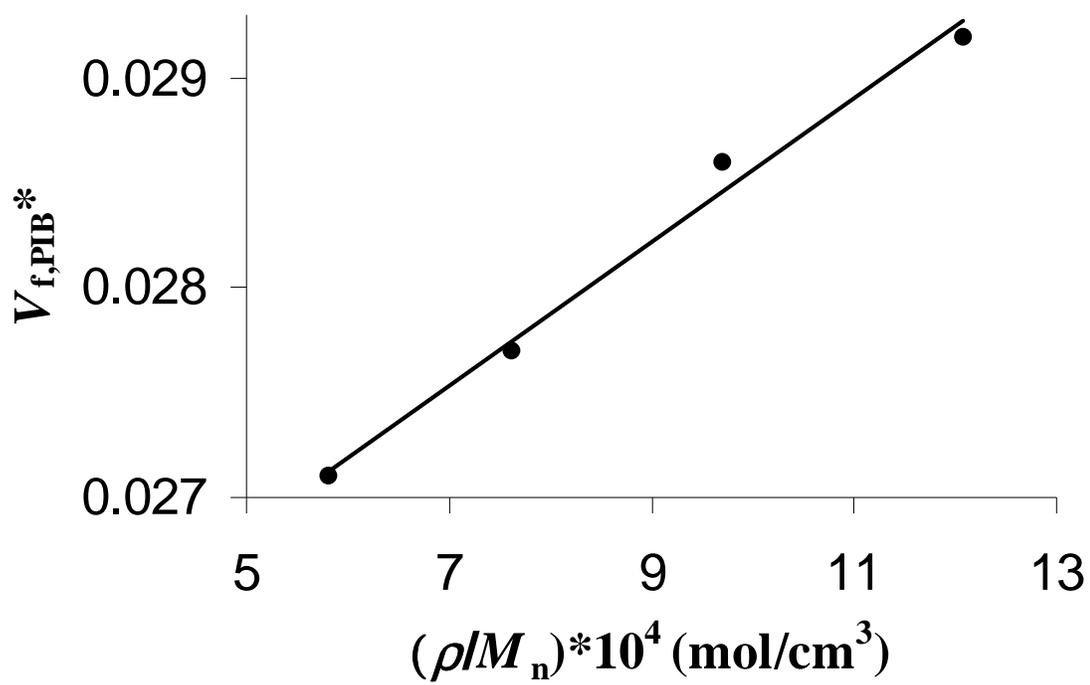

**Figure 8**

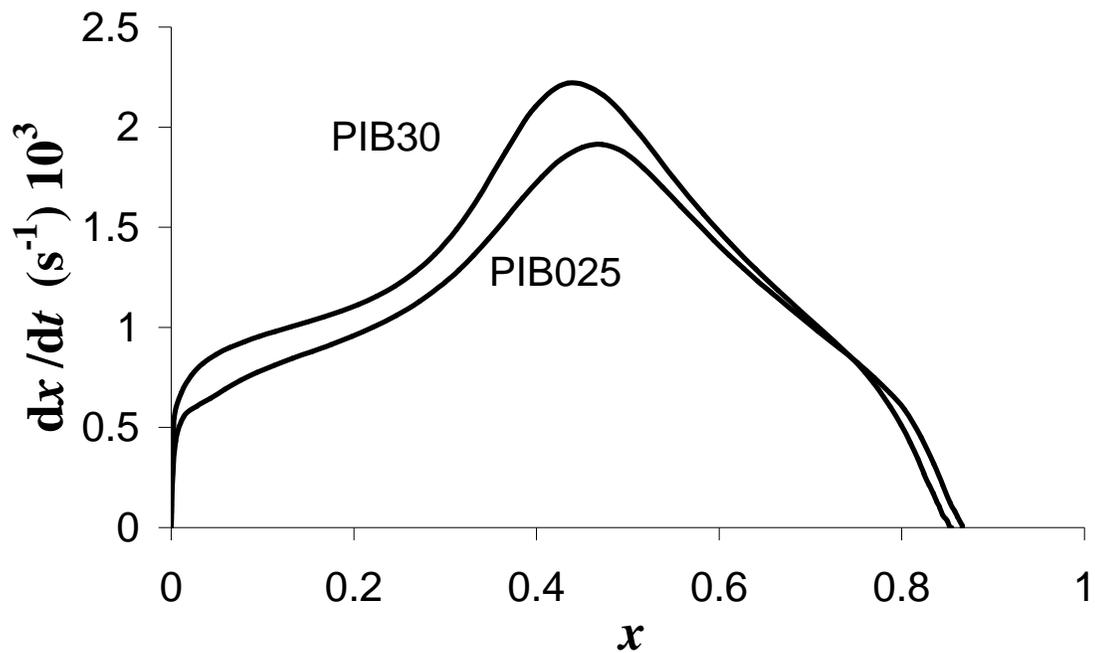

**(a)**

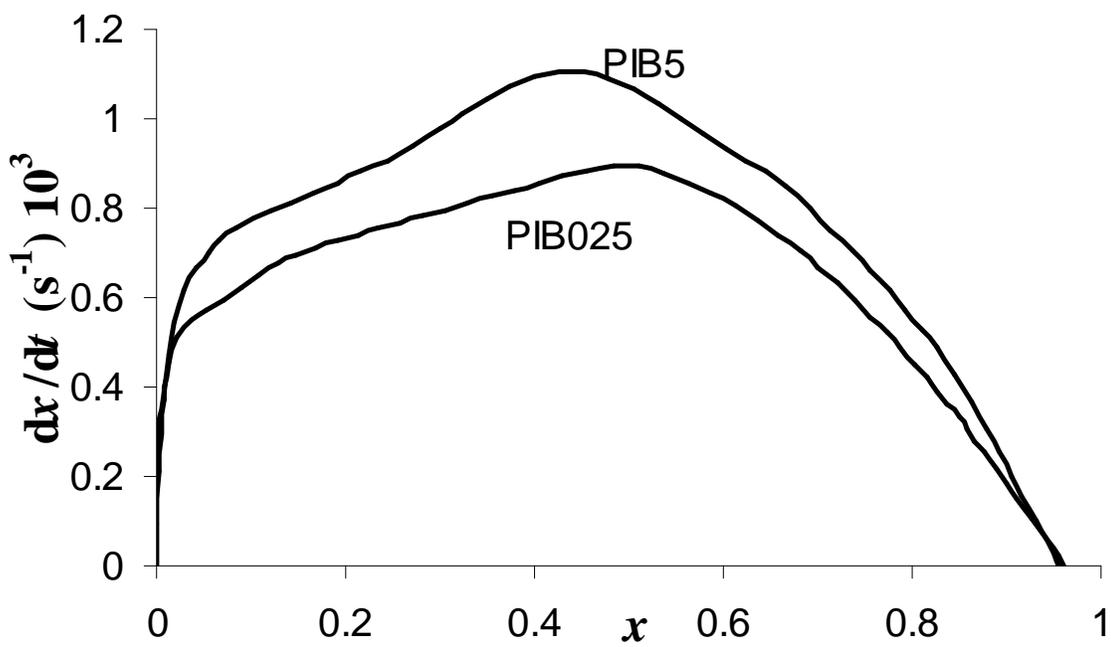

**(b)**

**Figure 9**

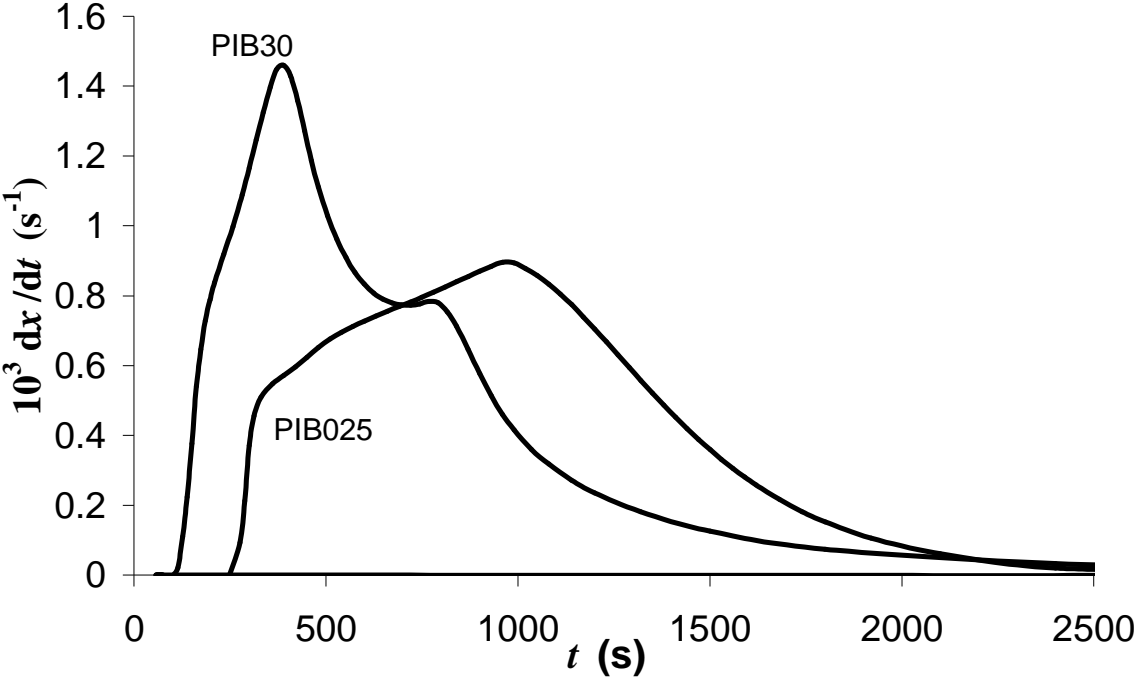

**Figure 10**

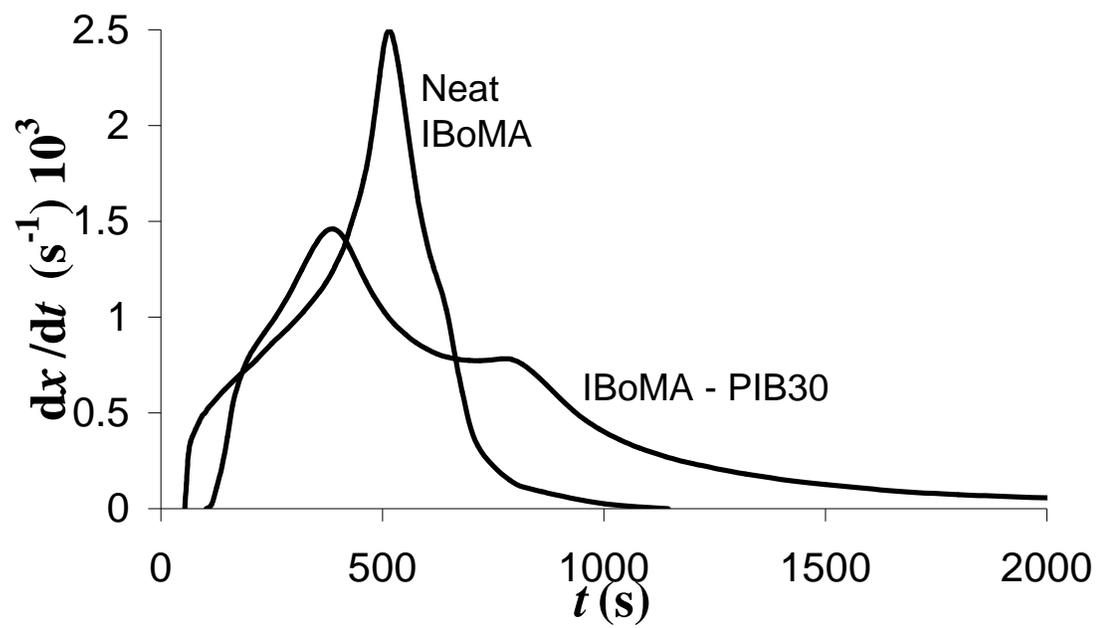